# A new methodology for comparing Google Scholar and Scopus

Henk F. Moed*, Judit Bar-Ilan** and Gali Halevi***

*Senior Scientific Advisor, Amsterdam, The Netherlands. Email: hf.moed@gmail.com

**Department of Information Science, Bar-Ilan University, Ramat Gan, 5290002, Israel. Email: Judit.Bar-Ilan@biu.ac.il

***The Levy Library, Icahn School of Medicine at Mount Sinai, One Gustave L. Levy Place, Mailstop 1102, New York, NY 10029, USA. Email: gali.halevi@mssm.edu

## Abstract

A new methodology is proposed for comparing Google Scholar (GS) with other citation indexes. It focuses on the coverage and citation impact of sources, indexing speed, and data quality, including the effect of duplicate citation counts. The method compares GS with Elsevier's Scopus, and is applied to a limited set of articles published in 12 journals from six subject fields, so that its findings cannot be generalized to all journals or fields. The study is exploratory, and hypothesis generating rather than hypothesis-testing. It confirms findings on source coverage and citation impact obtained in earlier studies. The ratio of GS over Scopus citation varies across subject fields between 1.0 and 4.0, while Open Access journals in the sample show higher ratios than their non-OA counterparts. The linear correlation between GS and Scopus citation counts at the article level is high: Pearson's R is in the range of 0.8-0.9. A median Scopus indexing delay of two months compared to GS is largely though not exclusively due to missing cited references in articles in press in Scopus. The effect of double citation counts in GS due to multiple citations with identical or substantially similar meta-data occurs in less than 2 per cent of cases. Pros and cons of article-based and what is termed as concept-based citation indexes are discussed.

## 1      Introduction

Google Scholar is increasingly used as a bibliometric tool to collect information on the citation impact of individual articles, researchers or scientific-scholarly journals, and competes with Thomson Reuters' Web of Science and Elsevier's Scopus. This paper compares Google Scholar and Scopus in terms of source coverage, citation impact of sources, citation counts to individual articles and their dependence upon 'double counts', indexing speed and data quality. Section 1.1 presents an concise review of the literature on the use of Google Scholar both as bibliographic and bibliometric tool, organized into three main themes – source coverage, citation impact and author-level studies – , while Section 1.2 gives an overview of the research questions addressed in the paper.

*1.1    Literature review*

*Source coverage*

The first area of study covered in this article is comparing Google Scholar coverage to that of Scopus. Earlier studies comparing Google Scholar to other scientific databases found significant gaps between its perceived and actual coverage. Jacsó (2005) noted the omission of highly relevant articles despite their availability in digital archives and Mayr & Walter (2007) discovered deficiencies in the coverage and up-to-datedness of the Google Scholar index when comparing international scientific journals from Thomson Scientific (SCI, SSCI, AH), open access journals and journals of the German social sciences literature database (SOLIS). More recent studies show significant improvement of Google Scholar coverage compared to its early years. Degraff, Degraff & Romesburg. (2013) demonstrated that the growth in the number of open-access journals and institutional repositories increases the number of articles readily available via Google Scholar in the area of geosciences.

Harzing (2013) examined Nobel Prize Winners in chemistry, economics, medicine and physics and their citations impact in Google Scholar, Scopus and Web of Science. She found that Google Scholar displays considerable stability over time and that coverage for disciplines that have traditionally been poorly represented in Google Scholar (chemistry and physics) is increasing rapidly. Lastly, in 2016 Harzing and Alakangas published the latest report of their longitudinal comparison between Web of Science, Scopus and Google Scholar. Examining 146 senior academics in five disciplines they found that the three databases display stable growth as far as the number of publications. However, the authors did find that Google Scholar still presents challenges especially in its inclusion of non-peer reviewed sources as citations, retrieval of duplicate, and thus redundant, documents in different versions which cause "stray citations" and it is possible to manipulate bibliometric indicators (Delgado López-Cózar, Robinson-García, & Torres-Salinas,2014). Our study examines the coverage levels of these sources by examining top cited papers in various disciplines and journals thus offering a wider perspective on the topic of coverage. In addition, this examination allows for an additional testing of previous findings and adds a new dimension of comparison to our knowledge.

*Citation impact*

By far the most studies comparing Google Scholar to other databases focus on citations counts. These studies examined Google Scholar coverage of citations to articles and authors (Bakkalbasi, Bauer, Glover & Wang, 2006; Neuhaus, Neuhaus, Asher & Wrede, 2006; Meho & Yang, 2007; Kousha & Thelwall, 2007; Kousha & Thelwall, 2008 ; Kulkarni, Aziz, Shams & Busse, 2009, Bornmann, Marx, Schier, Rahm, Thor, & Daniel, 2009; Levine-Clark & Gil, 2009; Mingers & Lipitakis, 2010; Bar-Ilan, 2010; Haddaway, Collins, Coughlin, & Kirk, 2015). The main findings of these studies revolve around two conclusions which are that Google scholar, much like Scopus and Web of Science, has certain coverage strengths in areas such as science and medicine but showed significant weaknesses in covering social sciences and humanities sources and demonstrated an English language bias, similarly to the other two databases. An interesting study by Haddaway et al. (2015) examined whether Google Scholar may replace commercial databases such as Web of Science and Scopus as a systematic review tool. The study found that the manner by which a user seeks content is significant. When searches are specific and the user is looking for particular artefacts, Google Scholar is able to retrieve them successfully. However, when more specific, complex searches are deployed, Google Scholar missed many of the important literature needed for a systematic review.

*Author level studies*

While Scopus and Web of Science produced compatible rankings for the studied authors, Google Scholar's rankings were significantly different mainly because of wider coverage of resources not indexed in the other two databases. While these resources generate more citations, it is difficult to predict rankings as Google Scholar does not have a clear indexing policy; an issue that has been pointed to in several other studies. The h-indices of highly cited researchers based on Google Scholar were considerably different from the values obtained using WOS or Scopus (Bar-Ilan, 2008). Similarly, h-index comparisons in the area of nursing was conducted by De Groote & Raszewski (2012) showed that Scopus, Web of Science, and Google Scholar provided different h-index ratings for authors and each database found unique and duplicate citing references again recommending that more than one tool should be used to calculate the h-index for nursing faculty because one tool alone cannot be relied on to provide a thorough assessment of a researcher's impact. This was recently confirmed by Wildgaard (2015) who also found that certain areas of science are better covered by Google Scholar and produce more favorable author rankings than others. The main recommendation in the study was for authors to be aware of the indexing coverage of each tool and not rely on one to compute their author-level impact indicators.

One of the features of Google Scholar is the listings of various versions per source when available. Many times the versions may contain a reference to the final published article on a publisher website which can be behind a paywall. Many times, however, these versions may contain pre-print versions in full text format. Citations to full text versions of articles on Google Scholar were studied by Jamali and Nabavi (2015). The study found that not only do full text versions found via ResearchGate or other educational repositories receive more citations but that there is a correlation between the number of full text versions found and the number of citations the article receives. Therefore, article h-index can show significant variations when measured by commercial databases and Google Scholar.

Many of the studies conducted extensive coverage and citations comparisons of Google Scholar to other databases and identified various degrees of differences between them. Overall it does appear that Google Scholar has improved its coverage over the years, especially in the Social Sciences yet its precision capabilities in terms of search are still lacking (Orduna-Malea, Ayllón, Martín-Martín & Delgado -López-Cózar, 2015). The lack of transparency with regards to its covered sources and inability to allow data exports for analysis presents difficulty in assessing its accuracy and usefulness as a source for evaluation metrics which involve citations counts (Ortega, 2014).

## 1.2     Research questions

### Source coverage

Perhaps the most striking feature of Google Scholar (GS) is that its citation counts are often so much higher than those generated in Web of Science or Scopus. Hence, the first research question of our study is: How does the coverage of GS compare to that of Scopus? More specifically: What is the ratio of a target article's number of citations retrieved from GS to its citation count obtained from Scopus? An additional question is how this ratio for target articles in Open Access (OA) journals compares to that of targets in non-OA periodicals. Other questions related to source coverage are: Which sources are indexed in Google but not in Scopus and vice versa, and which are covered by both? Focusing on the GS surplus, i.e., the citations in GS not found in Scopus, at which websites are their full texts available according to the web-link indicated in GS search results?

### Citation impact of sources

Coverage in terms of the volume of sources indexed is obviously an important aspect. But other aspects are very relevant as well. The first is their status measured in terms of citations. The research question addressed in this part of the study holds: how does the citation impact of documents in GS not indexed in Scopus (the GS-surplus) compare to that of documents both in GS and in Scopus, and to that of documents in Scopus that are not indexed in GS (the Scopus surplus). In this way, one obtains informetric data on the significance of the GS-and Scopus-surplus sources in terms of a core versus peripheral status in the written scholarly communication system, using Eugene Garfield's notions of citation indexing (Garfield, 1979).

### Statistical correlations

How good predictors are citation counts of individual articles generated in GS of citation rates obtained in Scopus and vice versa? If the two counts strongly correlate, it does not seem to matter much which of the two databases is used in an analysis of citation impact, at least of target articles published in sources indexed in both databases. Therefore, the research question addressed holds: how strong is the statistical correlation between citation counts at the level of individual articles between counts obtained in GS and those generated in Scopus?

### Indexing speed

A core issue in the current study is the speed of indexing. How up-to-date is a literature database? Can one find relevant documents published during the past week or month? Mayr & Walter (2007) report in 2007 that their tests show that Google Scholar is not able to present the most current data, but do not give details about these tests. In the current paper, indexing speed is studied by comparing the date at which documents published in Scopus-covered journals enter GS, compared to their entry date in Scopus itself.

### The effect of duplicates on citation counts

De Groote & Raszewski (2012) found "duplicate citing references" both in GS, WoS and Scopus, and also Adriaanse & Rensleigh (2013) observed that GS indexes multiple copies of the same article. Pitol and De Groote (2014) present an in-depth analysis of multiple versions in Google Scholar, and Valderrama-Zurián et al. (2015) on duplicate records in Scopus. Google Scholar often includes different versions of the same document. However, in search results, one particular version is displayed, while other versions are visible when clicking the "all versions" button. Since GS covers so many versions of a document, whereas Scopus indexes only the formally published, doi-ed version, users may fear that the surplus GS citation count of a particular article is at least partially caused by 'double counts', i.e., by multiple counting of the same citing document, available via different websites. Hence, a next research question is: to which extent do double counts occur in GS citation counts, and how does their frequency of occurrence compare to that in Scopus?

*Data quality and consistency*

This paper presents in Section 2 a series of important observations on GS data quality and consistency, especially relating to the internal consistency between the various database segments in GS, and to the accuracy of citation links.

### 1.3    Approach adopted in this paper and its limitations

The orientation of this article is primarily methodological. It proposes a series of methods of data collection, data handling and data analysis all aimed to provide insight into the differences in coverage between Google Scholar and Scopus. The methodology is applied to a set of 36 highly cited articles in 12 scientific-scholarly journals covering six subfields: political science and Chinese studies, two subfields from the social sciences and humanities; next, two subfields bridging social sciences & humanities and the formal sciences: computer linguistics, and library & Information science. Finally, two subfields were selected from the natural and life sciences: inorganic chemistry and virology.

The total number of analyzed citations to this set of 36 target documents amounts to about 7,000. The journals were selected in pairs, combining periodicals with distinct features in terms of country of the publisher (American versus European or Asian) and the journal's business model (Open Access (OA) versus non-OA). The study thus aims to reveal the variability of the differences between Google Scholar and Scopus across disciplines, journals, publisher country and access modalities, but does not allow a generalization expressing an overall difference of the two systems.

GS and Scopus data analyzed in the study were collected in the last week of July 2015. Google and Elsevier are continuously developing their products. As a result, the coverage of Google Scholar and Scopus change over time. The producers may reload their database, add new features to them, and correct errors. As a consequence, some results may already be out-of-date when this paper is published. Moreover, the effect of recently implemented features may not yet be fully visible.

The current article compares Google Scholar and Scopus in terms of source coverage indexing speed. It does *not* deal with the functionalities implemented in their online systems. Moreover, it will not give attention to specific bibliometric indicators such as h-index. Also, the study does *not* give a comprehensive analysis of the data quality and consistency in the two databases. But it *does* indicate several issues related to data quality and consistency of GS, but only in as far as encountered in the matching process of GS and Scopus documents.

## 1.4    Terminology and structure of the paper

In this article documents for which citation data are collected are denoted as *target* articles or in short as *targets*. The documents citing the targets are, depending upon the context, indicated as *citing documents* or as *citations*. The last term is used if the context focuses on citation *counts* or *number* of documents citing a particular set of targets, and the first if the analysis deals with other *properties* of the citing documents, especially those embodied in the various meta data fields. The outlets in which the documents are published (journals, books, conference proceedings) or the repositories in which they are posted, are denoted as *sources*.

Section 2 presents a list of the journals that were analyzed. It describes the processes of data collection and data handling, including the issue of duplicate documents and 'double counts' and data quality and consistency. The outcomes of the comparative analysis are presented in Section 3, ordered by research question. Finally, Section 4 contains a critical discussion of the outcomes. It focuses on the use of GS or Scopus for the calculation of indicators in quantitative research assessment, and presents a shortlist of pros and cons of Google Scholar for this type of use.

## 2    Data collection and data handling

### 2.1    Selection of target journals

Table 1 gives a list of the journals analyzed in the study, their publisher, number of articles published in 2014, principal countries of publishing authors, and access modality. Data were obtained from Scopus.com and from the Scopus Journal Title List of June 2015 (Scopus Journal Title List, 2015). The last two columns present two journal metrics: h5, the 5-year h-index of a journal for the time period 2010-2014, available in Google Scholar Metrics (https://scholar.google.com/intl/en/scholar/metrics.html ), and the Impact Per Paper (IPP) for the year 2014, available in Scopus, and defined as the average citation rate in 2014 of articles published in a journal during the three preceding years. The publication window of 2010-2014 was chosen because Google Scholar Metrics provided information for this period only at the time of data collection.

Table 1. Journals included in the analysis

| Nr | Subject Category in GS | Journal | Publisher | Nr Publ in 2014 (Scopus) | Principal author countries | Access modality | h5 2010-2014 | IPP 2014 |
|---|---|---|---|---|---|---|---|---|
| 1 | Chinese Studies | J Contemporary China | Routledge | 68 | 1. USA (15) 2. China (13) | SB | 23 | 1.32 |
| 2 | | China: An International Journal | Natl Univ Singapore | 28 | 1. China (12) 2. USA (6) | SB | 10 | 0.20 |
| 3 | Linguistics / Computer Science | Computational Linguistics | MIT Press Journals | 37 | 1.USA (14) 2. UK (11) | OA | 31 | 2.42 |
| 4 | | Computer Speech & Language | Academic Press | 108 | 1.USA (22) 2. Spain, UK (14) | SB | 32 | 1.72 |
| 5 | Inorganic Chemistry | Inorganic Chemistry | Am Chem Soc | 1,518 | 1.USA (454) 2.China (303) | SB | 77 | 4.58 |
| 6 | | Eur J Inorg Chem | Wiley-VCH Verlag | 776 | 1.Germany (175) 2.China (114) | SB | 34 | 2.61 |
| 7 | Libr & Inf Sci | Scientometrics | Springer/ Akadémiai Kiadó | 402 | 1.China (90) 2. Spain (55) | SB | 46 | 2.13 |
| 8 | | D-Lib Magazine | Corp. Natl. Research Initiatives | 55 | 1.USA (20) 2.UK (9) | OA | 18 | 0.85 |
| 9 | Political Sci | Am J Political Sci | Wiley-Blackwell | 65 | 1.USA (52) 2.Israel, Switzerl., UK (3) | SB | 58 | 3.34 |
| 10 | | Eur J Political Res | Wiley-Blackwell | 47 | 1 UK (12) 2.USA (10) | SB | 36 | 1.79 |
| 11 | Virology | J Virology | Amer Soc Microbiol | 1,312 | 1.USA(789) 2.China(142) | SB/ OA after 6 months | 88 | 4.32 |
| 12 | | PLoS Pathogens | Public Lib Science | 693 | 1.USA(385) 2.UK(102) | OA | 99 | 7.22 |

Legend to Table 1: Access modality: SB = Subscription Based; OA = Open Access, i.e., based on Author Pay business model, and/or freely available.

To be able to carry out the data collection and analysis methods explored in the current study, journals were selected from GS-Metrics, which provides citation data for a limited set of relatively highly cited journals publishing almost exclusively in English, and is therefore biased in terms of journal impact, country of publisher and publication language. The selection of GS-Metrics journals in the study aims to cover sources from a range of disciplines (humanities, social sciences, natural science, computer science, and social sciences), with distinct geographical distribution of authors – selecting both European and US dominated journals and an Asian periodical as well – and access modalities - fully Open Access versus subscription based journals. The classification OA-non-OA does not take into account articles in subscription-based journals that are made OA by selecting an "open choice" option.

## 2.2    Data collection and handling

Figure 1 gives a schematic overview of the various steps in the data collection and handing process, while Table 2 presents a list of the datasets that were created in this process. Focusing on information about the *target* articles, per target journal three datasets were created: a set of the 200 most frequently cited target documents in GS Search (SET 1), a set of documents available via GS Metrics (SET 5) and a set of the 200 most frequently cited documents in Scopus (SET 7). These three datasets were linked to one another using a match-key based on words from the document titles and the author lists (for details, see Section 2 below). These sets were considered sufficiently large to compare for each target journal the *upper part* of the citation distributions at the article level extracted from the two databases. The sets of 200 most often cited documents in GS-Search and in Scopus did not fully overlap. Moreover, for some journals the total number of published target documents was below 200. To deal with these limitations, Sections 3.1 (analysis of source coverage) and 3.5 (statistical correlations) present an analysis of a set of the 100 most frequently cited documents in GS Search (from SET 1) which were also among the 200 most often cited articles in Scopus (from SET 7). Both for GS-Search and for GS-Metrics the publication window was 2010-2014. But the end date of the citation window applied in the GS-Search counts is about one month later than that for the GS-Metrics counts: July 2015 versus June 2015.

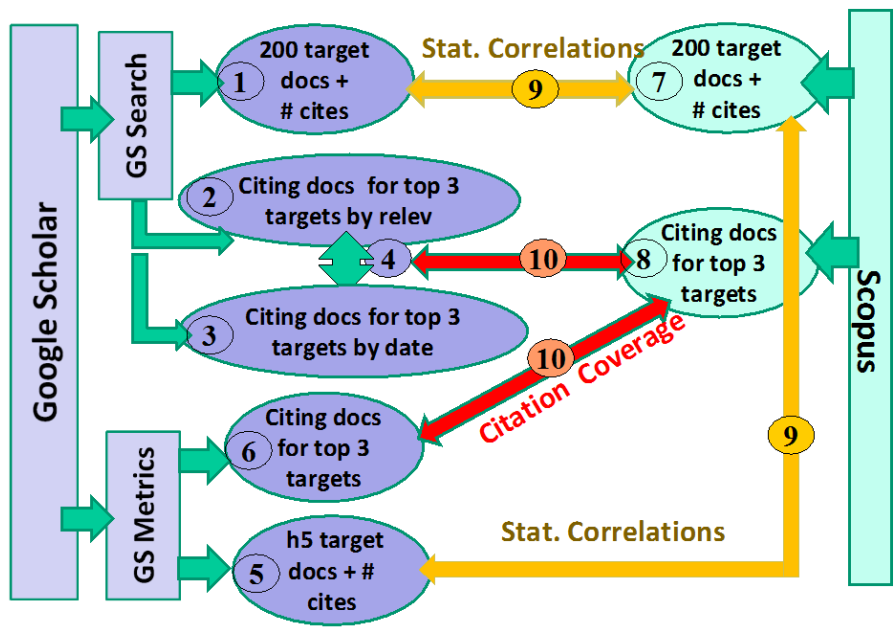

Figure 1: Process of data collection per journal

Table 2: Datasets created

| Set or Step no. | Dataset or Step |
|---|---|
| 1 | Extract via the Advanced Search option in Google Scholar (denoted as GS Search) all documents published in a given journal during the time period 2010-2014, and extract the 200 most frequently cited documents (SET 1). |
| | Select from SET 1 the three most frequently cited documents. These are the target documents in the citation analysis. This set is defined as the TARGET TOP 3 SET. |
| 2 | Extract for each of document in the TARGET TOP 3 SET all documents in GS Search citing a particular target and published during the time period from 2010 up to date (July 2015), sorted by "relevance" (SET 2). |
| 3 | Extract for each target document in the TARGET TOP 3 SET all documents citing a particular target and processed for GS Search during the past 365 days, by sorting the list generated in STEP 3 "by date" (SET 4). |
| 4 | Combine SET 2 and SET 3 so that the combined set (SET 4) contains for each document information on the entry date in GS (as far as available) |
| 5 | Extract via the "Metrics" module in Google Scholar (denoted as GS-Metrics) for a given journal all documents listed in this module, i.e., cited at least h5 times (SET 5). h5 is the value of the 5-year Hirsch Index for the journal. |
| 6 | Extract via the "Metrics" module in Google Scholar for each target document included in SET 2 all documents citing a particular target (SET 6). |
| 7 | Select from Scopus.com all documents published in a given journal during the time period 2010-2014, and extract the 200 most frequently cited documents (SET 7), |
| 8 | Extract for each target document in the TARGET TOP 3 SET all documents indexed in Scopus.com citing a particular target and published during the time period from 2010 up to date (July 2015) (SET 8). |
| 9 | Match-merge SETS 1, 5 and 7 at the level of individual documents. The resulting dataset is denoted as "ALL TARGETS". It contains for each target document citation counts extracted from Google Scholar Search, counts from Scopus.com, and – for the h5 most frequently cited documents – citation counts from Google Scholar Metrics. |
| 10 | Match-merge SETS 4, 6 and 8 by target article at the level of individual citing documents. The dataset created in this way is labeled "ALL CITATIONS". It combines for each citing document the web domain via which it is available (in GS Search) with information on the date at which it was indexed in GS (in GS Search, for docs indexed during the past 365 days only), with source information available from GS Metrics (especially its journal/source title) and, whenever a match was found between a GS and a Scopus citing document, with source information from Scopus. |

A dataset denoted as TARGET TOP 3 SET was created containing the three most frequently cited documents in GS Search published in the study set of 12 journals listed in Table 1. About 72 per cent of these articles were published in 2010, and 16 per cent in 2011. This bias towards the oldest articles in the set is caused by the fact that GS-metrics counts citations during a fixed time period (2010-2014), so that 2010-papers can be followed during at least 4 years, but 2014-papers for at most one year. The above mentioned sets per journal of the 100 most highly cited documents in GS Search reveals the same bias, though less pronounced: 40 per cent of documents were published in 2010, and 25 per cent in 2011.

Next, four datasets were collected and combined with detailed information about the documents citing the articles in the TARGET TOP 3 SET. Three datasets relate to GS, and one to Scopus. The first is the list of citing documents sorted by "relevance" and obtained by clicking in a Google Search result on the number of citations of the target articles analyzed (SET 2). In this way, for each (citing) document information is obtained on the document title, the first part of the author list, the first part of the source title, and the "preferred" web domain to the website via which the full text can be retrieved. Sorting this list on line "by date", extracting the document records (SET 3) and match-merging them with SET 1 into SET 4, an additional piece of information was added , namely, the time elapsed at the date of data collection from the moment it was indexed in GS. This information is only available for documents indexed during the previous 365 days.

Next, information on documents in the TARGET TOP 3 SET was extracted from Google Metrics (SET 6). These records contain a much longer part of the source title of the citing document. By combining SETS 4 and 6, a compound dataset was created containing for each document in GS-Search information on document title, web domain, the number of citations in GS-Search, the time elapsed since its entry date in GS (for documents indexed during the previous 365 days only) and, in as far as available, from GS-Metrics, a more complete source title, publication year, volume and starting page number, as well as the number of citations in GS-Metrics. In a next step, this compound GS dataset was matched against the citation datasets extracted from Scopus (SET 8).

Documents with meta data in non-Latin characters, especially those in Chinese and Russian, were deleted. The algorithm also deleted two records of which the title did not contain any word longer than 3 characters and complete source data. In all, the raw dataset of 11,367 documents extracted from GS-Search, including both target documents and citations, 230 records (2.0 %) were deleted. From the 7,424 documents downloaded from GS Metrics, 183 (2.5 %) were deleted. From the 7,424 documents downloaded from GS Metrics, 183 (2.5 %) were deleted. None of the 5,967 documents extracted from Scopus were deleted. Diacritic characters in the data, containing accents such as à, á, ñ, were resolved to their base form (a, a, n, respectively). All data were extracted during the time period between 22 July and 31 July, 2015. For each journal, Scopus and Google Scholar data were downloaded on the very same day. In the collection of citation data, citing documents marked as [Citation] in GS were included. Such documents are extracted from a cited reference list of a source document, rather than being indexed as a source document itself. Of the 6,536 citations in citing documents extracted from Google Scholar, 2.5 per cent are marked as [Citation].

## 2.3    Match-merging and duplicates

An analysis of matching and duplicate records was conducted as follows. Four match-keys were defined, listed in Table 3. Title words were extracted from a document's full title using a set of separators such as space, comma, quotes and brackets, and selecting words with at least 4 characters. Author names from the various databases were first converted into a standard format. It is noteworthy that the publication year is not included in any of these match-keys. The reason is that the publication year is an ambiguous concept, as it may refer either to a document's online publication date or to the issue date. The records extracted via Google Search contain the online year, whereas those in Scopus, but also the records from the "frozen" GS Metrics dataset tend to contain the issue date. As can be seen in Table 3, most records were matched using the full publication key.

Table 3: Match-keys applied in match-merging Google Scholar and Scopus

| | Match-key | Details | % matched targets GS-Scopus (n=1,894) | % matched citations GS-Scopus (n=3,246) |
|---|---|---|---|---|
| 1 | Full publication key | The first 6 characters of the first author's last name, plus the first 10 words from the title with a word length of at least 4 characters | 92.7 % | 90.2 % |
| 2 | Title key | The title-based part of the full publication key (i.e., author name part is deleted from full publication key) | 3.9% | 2.0 % |
| 3 | Short publication key | The first 6 characters of the first author's last name, plus the first title word. | 3.2 % | 6.9 % |
| 4 | Source-based key | The first 6 characters of the first author's last name, plus the volume number and starting page number | 0.2 % | 0.9 % |

In each of the three data files containing GS Search, GS Metrics, and Scopus records, respectively, candidate-duplicates were identified, consecutively applying the four match-keys. Two citing documents could be candidate-duplicates only if they are citing *the same* target article. Pairs were formed of documents with the same value of a particular match-key, and all available data fields were compared to categorize them according to the degree of overlap between their elements, in terms of 'being identical', 'showing a large degree of similarity', or 'showing a low degree of similarity'. A major indexing problem is how to identify duplicate records if their meta data are written in different languages. The method applied in the current study is partially capable of identifying such case, namely by applying the source-based key defined in Table 3. The percentage of duplicate pairs was 4% for GS, 5% for GSM and 2% for Scopus. In the analyses presented below, duplicate documents showing a large degree of similarity or being identical were deleted from the data files. More details can be found in Appendix A1.

*2.4    Data quality and consistency*

In the data collection process outlined in Section 2, the following observations were made on the data quality and consistency of the Google Scholar data. When match-merging the two downloaded citation datasets sorted "by relevance" (SET 2 in Figure 1) and "by date" (SET 3), respectively, 4.5 per cent of records in the second were not found in the first. Also, 4 per cent of the about 550 target articles extracted from GS Metrics (SET 5) could not be found in GS Search (SET 1), and 2.5 per cent of the about 6,700 citations in GS-Metrics (SET 6) could not be found in the set of GS-Search citations (SET 2). Finally, one of the three most frequently cited targets articles in Journal of Virology is cited in GS 270 times, but a secondary analysis revealed that 180 of these were linked erroneously to this target article, all extracted from a particular (Brazilian) journal available via a Cuban website.

## 3. **Results**

### 3.1    *Source coverage: numbers*

Table 4 presents per target journal, for the 100 most frequently cited articles in GS Search for which citation counts in Scopus were available in the study (i.e., which were among the top 200 cited articles in Scopus, see Section 2.1), two measures of the ratio of Google Scholar over Scopus citations. The first is a 'globalized' ratio, which is defined as the sum of citations in GS to the 100 targets divided by the same sum of citations in Scopus. The second is an averaged ratio, calculated per journal as the mean over all its 100 target articles of the ratio of GS over Scopus citations at the level of an individual article. In case the citation count in Scopus was 0, which happened in 3 per cent of the cases, it was set to a value of one. Unless specified otherwise, in the comparison between Google Scholar and Scopus, the Google Scholar set is formed by merging the GS Search and the GS Metrics subsets. In this way, the combined set, denoted as "the Google Scholar" set, contains 214 records in GS Search not found in GS Metrics (3.3 per cent), and 117 records in GS Metrics not found in GS Search (1.8 per cent).

Table 4: Ratio of Google Scholar over Scopus citations for the top 100 articles in 12 target journals

| field | Target_journal | Total Target articles* | Sum cites in GS* | Sum Cites in Scopus* | Globalized Ratio GS / Scopus Cites | Averaged Ratio GS / Scopus Cites |
|---|---|---|---|---|---|---|
| *ALL* | *ALL* | 1200 | 67,785 | 43,732 | 1.6 | 2.4 |
| Chinese Stud | China: An International Journal | 100 | 330 | 118 | 2.8 | 1.7 |
| Chinese Stud | J Contemporary China | 100 | 2,006 | 973 | 2.1 | 2.6 |
| Comput Ling | Computational Linguistics | 100 | 3,732 | 1,238 | 3.0 | 4.1 |
| Comput Ling | Computer Speech & Language | 100 | 3,336 | 1,625 | 2.1 | 2.7 |
| Inorg Chem | European J Inorganic Chemistry | 100 | 3,717 | 4,643 | 0.8 | 0.8 |
| Inorg Chem | Inorganic Chemistry | 100 | 10,245 | 9,440 | 1.1 | 1.1 |
| Libr & Inf Sci | D-Lib Magazine | 100 | 1,152 | 403 | 2.9 | 3.4 |
| Libr & Inf Sci | Scientometrics | 100 | 5,356 | 2,683 | 2.0 | 2.1 |
| Polit Sci | Am J Political Sci | 100 | 8,661 | 2,641 | 3.3 | 3.9 |
| Polit Sci | Eur J Political Res | 100 | 3,710 | 1,264 | 2.9 | 3.5 |
| Virology | Journal of Virology | 100 | 11,809 | 8,812 | 1.3 | 1.4 |
| Virology | PLoS Pathogens | 100 | 13,731 | 9,892 | 1.4 | 1.4 |

* Publication window 2010-2014, citation window 2010-June/July 2015

Table 4 shows that for almost journals, and especially for the aggregate of all targets, the globalized ratio is lower than the averaged one (1.6 against 2.4). This is because the ratio of GS over Scopus citations of an article tends to decline as its number of citations in Scopus increases. In fact, these to variables show a Pearson correlation coefficient of -0.35, which is significant at the 99 per cent confidence level.

Table 5. Overlap in citations between Google Scholar (GS) and Scopus (SC) per target journal

| Field | Target journal | Total # Cites in GS* | Total # Cites in SC* | # Cites both in GS & SC* | Total # Unique Cites | Ratio GS/SC Cites | % Cites in GS out of Total # Unique Cites | % Cites in SC out of Total # Unique Cites | Stdev/ Mean Ratio GS/SC Cites |
|---|---|---|---|---|---|---|---|---|---|
| *ALL* | *ALL* | 6,536 | 3,651 | 3,246 | 6,941 | 1.8 | 94.2 | 52.6 | . |
| Chinese Studies | China: An Internat. Jrnl | 49 | 25 | 17 | 57 | 2.0 | 86.0 | 43.9 | 7.0 |
| | J Contemporary China | 248 | 137 | 77 | 308 | 1.8 | 80.5 | 44.5 | 18.2 |
| Comput Linguist | Computational Linguistics | 1,008 | 401 | 368 | 1,041 | 2.5 | 96.8 | 38.5 | 24.2 |
| | Computer Speech & Lang | 479 | 238 | 217 | 500 | 2.0 | 95.8 | 47.6 | 24.1 |
| Inorg Chem | Eur J Inorganic Chem | 344 | 366 | 253 | 457 | 0.9 | 75.3 | 80.1 | 39.4 |
| | Inorganic Chemistry | 817 | 707 | 664 | 860 | 1.2 | 95.0 | 82.2 | 6.8 |
| Libr & Inf Sci | D-Lib Mag | 171 | 60 | 49 | 182 | 2.9 | 94.0 | 33 | 29.8 |
| | Scientometrics | 663 | 324 | 294 | 693 | 2.0 | 95.7 | 46.8 | 6.4 |
| Political Sci | Am J Political Sci | 974 | 319 | 286 | 1,007 | 3.1 | 96.7 | 31.7 | 11.8 |
| | Eur J Political Res | 485 | 145 | 136 | 494 | 3.3 | 98.2 | 29.4 | 9.3 |
| Virology | J Virology | 534 | 413 | 406 | 541 | 1.3 | 98.7 | 76.3 | 6.2 |
| | PLoS Pathogens | 764 | 516 | 479 | 801 | 1.5 | 95.4 | 64.4 | 9.8 |

* Publication window 2010-2014, citation window 2010-June/July 2015

Table 5 compares for the Target Top 3 Set in each of the 12 target journals the number of citations obtained in Google Scholar (GS) and Scopus, and the overlap between these two databases, i.e., the number of citing documents indexed in both databases. Journals are arranged by subject field. It shows for instance that the number of citations found in GS to the three top target articles in the *American Journal of Political Research* is 3.1 times the number of cites to these papers indexed in Scopus, but *European Journal of Inorganic Chemistry* it is 0.9. For all journals combined, the ratio of GS and Scopus citations amounts to 1.8. Within the set of citations in GS, the ratio of the number of citations in GS indexed in GS only over that of citations found both in GS and in Scopus amounts to 1.0 ((6,536-3,246)/3,246). The latter ratio is further discussed in Section 3.6 near Figure 5. The last column provides insight into the amount of variability among target articles.

Table 6 presents the distribution of the publication years of the 6,536 citing documents in GS and 3,246 documents in Scopus analyzed in Table 5. For documents in GS the publication year in GS-Search was taken. The table shows that for 11.4 per cent of citing documents in GS the publication year is unavailable. This is mainly due to documents for which no publication year is available in the source in which it is deposited. The GS Search year indicates the online year rather than the formal publication year. During 2010-2014, the annual percentages increase. This is due to the fact that the target articles are published during 2010-2014, and since it takes several years for citation impact to mature. Also, the number of citable documents increases during these years. As citations were counted from 2010 up until July 2015 the year 2015 is incomplete. This is why the percentages drop so sharply in 2015.

Table 6: Distribution of publication years of citing documents in GS and Scopus in percentages

| Publication years of citing documents | Google Scholar (n=6,536) | Scopus (n=3,246) |
| --- | --- | --- |
| N.A. | 11.4 | 0.0 |
| <=2007 | 0.4 | 0.0 |
| 2008 | 0.4 | 0.0 |
| 2009 | 0.8 | 0.0 |
| 2010 | 3.4 | 2.6 |
| 2011 | 8.5 | 9.6 |
| 2012 | 13.5 | 15.8 |
| 2013 | 19.2 | 21.8 |
| 2014 | 26.0 | 31.5 |
| 2015 | 16.3 | 18.5 |

*3.2 Further specification of the overlap*

Citing documents in Google Scholar that were *not* matched to a corresponding citing document in Scopus– were subdivided into two sub-categories: "in GS only but published in a journal indexed in Scopus", and "in GS only and *not* published in a journal indexed in Scopus". In order to determine whether or not a journal was indexed in Scopus, two approaches were adopted. Firstly, the journal title was matched against the list of *active* journals in the Scopus Journal Title List for June 2015. The second approach made use of the fact that if a document in GS is correctly matched to a corresponding document in Scopus (in this process the source titles do not play a role), at the same time a pair of corresponding source titles is created. After manual checks, keeping only correct matches, all GS documents published in the thus identified sources were earmarked, and added to the set of documents published in Scopus covered sources (if they were not yet included). The first approach focuses on journals and ignores book titles and conference proceeding sources. Although the second approach does not suffer from this limitation, standardization of book and conference proceedings title are not as strictly standardized as are journal titles; hence, some of the books or proceedings indexed in Scopus many not have been properly identified.

In a similar manner, documents categorized as "In Scopus only" were categorized into the sub-categories "In Scopus only but in source indexed in GS", and "In Scopus only and source *not* indexed in GS". This sub-categorization is more difficult to make than the one related to sources "in GS only", as there is no full thesaurus available of sources indexed in GS. Hence, in this case, only the second approach could be applied, i.e., matching GS source titles against a list of GS sources matched to at least one source indexed in Scopus. Table 7 shows a breakdown of citing documents across the various sub-categories.

Scopus indexes so called *articles in press* (AIP). These are articles are in the publication process, and have not yet been formally published in a journal issue, but they have been published online on a publisher's website. As a rule, in Scopus the cited reference lists of articles in press are *not* indexed. Hence, if these articles cite one or more of the 36 target articles in the study set, they could *not* be retrieved in a citation search in Scopus to these targets. As a consequence, the sub-category 'citations in GS only but published in a source indexed in Scopus ' consists of a certain fraction of citations listed in cited reference lists in articles-in-press the meta data of which are indexed in Scopus, but the cited reference lists are not.

To give at least some indication of the value of this fraction, the following approach was adopted. In a first step, a list was created of the publishers of documents in the subcategory 'citations in GS only but published in a source indexed in Scopus'. Thirteen 'big' publishers were identified. Next, for these publishers it was examined on 8 Dec. 2015 whether they had published any articles indexed in Scopus as AIP in 2015. For Elsevier, Wiley-Blackwell (except for Journal of the Association of Information Science and Technology), Springer, Cambridge University Press, Taylor & Francis, Akadémiai Kiadó, OUP, and Wolters Kluwer Health , articles in press were found. For SAGE, Macmillan, RSC and ACS no AIP were found.

Table 7. Categorization of citing documents

| Citing documents sub-category | Sub-subcategory | Frequency | Per cent |
|---|---|---|---|
| Both in Google Scholar (GS) and in Scopus | | 3,246 | 46.8 % |
| in GS only but published in a source indexed in Scopus | | 555 | 8.0 % |
| | Publisher has Articles in Press (AIP) in Scopus in 2015 | 271 | 3.9 % |
| | Publisher has no AIP in Scopus in 2015 | 284 | 4.1 % |
| in GS only and not published in a source indexed in Scopus | | 2,735 | 39.4 % |
| In Scopus only but in source indexed in GS | | 227 | 3.3 % |
| In Scopus only and not in source indexed in GS | | 178 | 2.6 % |
| Total | | 6,941 | 100 % |

Five of the several dozens of remaining, smaller publishers were checked and no AIP were found. If one assumes that none of these smaller publishers has AIP in Scopus, this analysis suggest around half of the citations in GS not found in Scopus but published in Scopus-indexed journals may by contained in source articles indexed in Scopus as AIP. This percentage represents an upper bound, as it assumes that all documents in journals "with AIP" in Scopus are actually articles in press. A follow-up study should analyze this issue in more detail. Figure 2 shows per journal the distribution of the citing documents across subcategories. Journals are arranged by field, as in Table 6. It shows for instance that *European Journal of Inorganic Chemistry* has by far the largest percentage of citations published in GS covered sources but not found in GS.

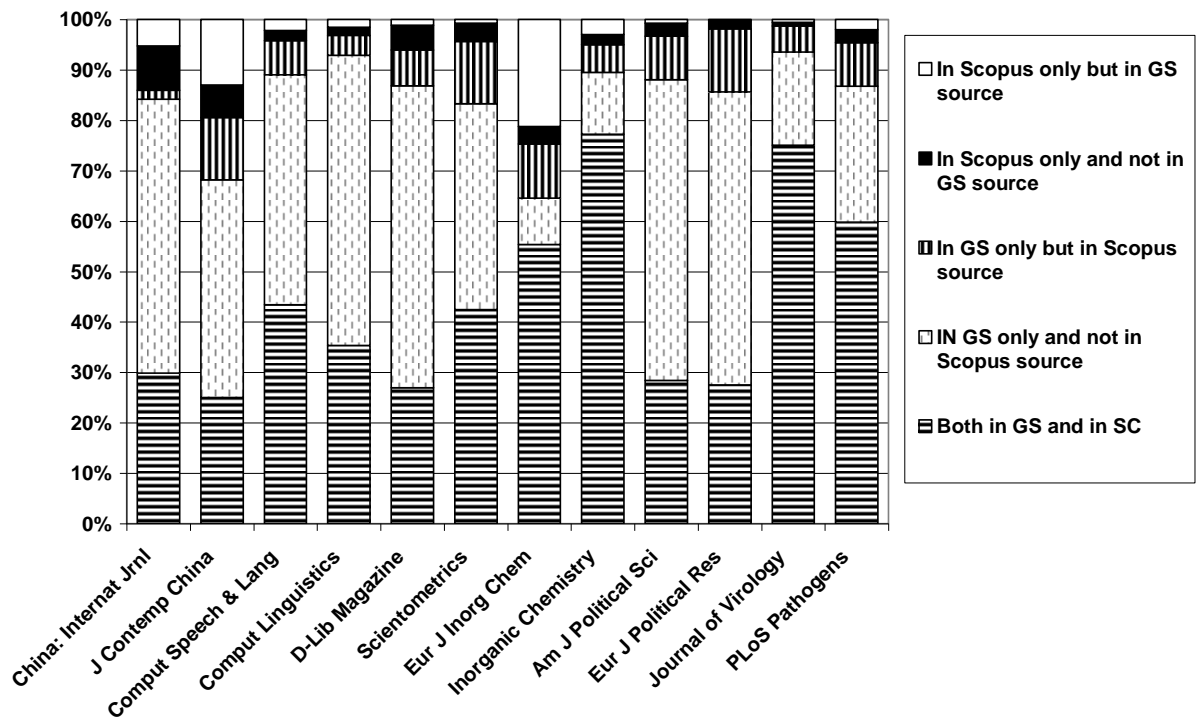

Figure 2. Distribution per journal of the citing documents across subcategories. The order of the subcategories in the charts from top to bottom is the same as that of the sub-category names at the right-hand side of the figure.

### 3.3    Degree of overlap: sources and web domains

The distribution of documents across sources in Google Scholar – journals, books, conference proceedings, but also repositories, archives – is highly skewed. Table 8 presents statistics of three distributions: the distribution of citations among sources in Google Scholar not indexed in Scopus, and among web domains in GS Search of sources not indexed in Scopus, and the distribution of citations among sources in Scopus not linked to a source title in Google Scholar Metrics. Table 8 shows the skewness of the distribution of citing documents across sources in GS and Scopus and web domains in GS.

The web domains appearing 50 times or more in the set of unique GS citations are: Google Books (156), Springer (140), SSRN (93), Researchgate (86), the ACM Digital Library (63), Arxiv (54) and ACL (53). More details on the source distribution of both GS only sources and sources in Scopus not found in GS can be found in appendix A2.

Table 8. Distribution of (citing) documents among web domains and sources in GS and sources in Scopus

| Document Sub-Universe | Entities | # Docs | # (%) Docs with unidentified entities | Total # entities | # (%) entities appearing once | Maximum number of appearances of a specific entity |
|---|---|---|---|---|---|---|
| Docs in GS not published in Scopus sources | Web domains | 2,735 | 176 (6.%) | 1,082 | 795 (73 %) | 156 |
| | Sources | 2,735 | 1,033 (38 %) | 1,214 | 999 (82 %) | 53 |
| Docs in Scopus not linked to a source in GS | Sources | 178 | 0 (0 %) | 149 | 130 (87 %) | 4 |

Table 8 presents for two sub-universes of documents – those in GS not published in Scopus sources and those in Scopus not linked to a GS source – information on the distribution of documents among web domains and sources. In the first subset the total number of documents amounts to 2,735. For 6 per cent of these there is no information on the document's web domain, while for 38 per cent the source title is missing. It must be noted that source titles were obtained from GS-Metrics. Apparently the bibliographic information on (citing) documents in GS-Metrics is rather incomplete. The total number of different web domains is 1,082. 73 per cent of these occur only once, i.e., are assigned to one single document only, while the most frequently occurring domain has 156 appearances. For source titles the distribution is even more skewed: 82 per cent of titles appears only once, and the maximum count is 53. The number of documents in the second sub-universe – docs in Scopus not linked to a GS source – is much lower than that in the first, namely 178. Almost 90 per cent of the 130 sources occurs only once.

## 3.4    Citation impact of sources

Table 9 gives information on the citation impact of the *documents which cited the target documents* analyzed in the study. It presents the average, age-normalized citation rate of the various types of citing documents retrieved from Google Scholar and Scopus, respectively. The age-normalized citation rate corrects for differences in publication years of the citing documents and was calculated in each of the two databases separately by dividing the number of citations to a (citing) document published in a particular year by the average citation rate of all (citing) documents published in that year. In this way, the average normalized citation rate across all (citing) documents (from all years) in each database amounts exactly to 1.0, but direct cross-database comparisons cannot be made.

The age normalization applied in this study is a first approximation; more advanced age normalization is feasible, accounting for differences among subject fields. But in the current study, with its methodological focus, the results properly indicate orders of magnitude.

*From the GS Search perspective*, the citation impact in GS of documents in GS *not* published in Scopus-covered sources is 79 per cent lower (100*(1.49-0.31)/1.49) than that of documents indexed in both databases. *From the Scopus perspective*, the impact in Scopus of documents in Scopus *not* published in GS sources is 86 per cent lower than that of sources covered in both. According to a Tukey test both differences are statistically significant at the 99 per cent confidence level. This is also true for the impact difference in GS surplus documents between the impact of documents in sources indexed in Scopus and that in non-Scopus covered sources (0.76 versus 0.31).

Table 9. Differences in age normalized citation rates between types of documents from Google Scholar Search and Scopus perspective

| Type of document | From Google Scholar Search perspective | | From Scopus perspective | |
|---|---|---|---|---|
| | No. Docs (2010-July 2015) | Average Normalized Citation Rate | No. Docs (2010-July 2015) | Average Age-Normalized Citation Rate |
| Both in GS and in Scopus | 3,145* | 1.49 | 3,238 | 1.03 |
| In GS only and not in Scopus source | 2,049 | 0.31 | 0 | . |
| In GS only but in Scopus source | 494 | 0.76 | 0 | . |
| In Scopus only and not in GS source | 0 | . | 171 | 0.14 |
| In Scopus only but in GS source | 0 | . | 227 | 1.16 |

*93 (citing) documents that were extracted from GS Metrics, and not included in GS Search results are not included

## 3.5    Statistical correlations

Table 10 gives the Pearson and Spearman coefficients of the correlation between GS Search and Scopus citation counts and between GS Metrics and Scopus counts at the article level. The first is based on the set of the 100 most frequently cited documents in GS Search for which Scopus citation counts were available in the study (i.e., which were among the top 200 in Scopus), and the second on a subset of the above set of documents for which citation counts are available in GS Metrics, i.e., with counts up or above the journal's value of h5. Figures 3 and 4 present scatter plots for the two journals with the highest and the lowest value of the Spearman correlation coefficient: *Inorganic Chemistry* and *Scientometrics*.

Table 10. Linear and rank correlation coefficients between GS and Scopus citation counts at the article level

| Field | Target Journal | GS Search-Scopus | | | GS Metrics-Scopus | | |
|---|---|---|---|---|---|---|---|
| | | N | Pearson | Spearman | N | Pearson | Spearman |
| Chinese Stud | China: An International Journal | 100 | 0.87 | 0.77 | 10 | 0.84 | 0.74 |
| | J Contemporary China | 100 | 0.92 | 0.71 | 23 | 0.96 | 0.81 |
| Computat Ling | Computational Linguistics | 100 | 0.96 | 0.83 | 31 | 0.99 | 0.85 |
| | Computer Speech & Language | 100 | 0.93 | 0.83 | 32 | 0.88 | 0.68 |
| Inorg Chem | European J Inorg Chemistry | 100 | 0.89 | 0.82 | 34 | 0.81 | 0.77 |
| | Inorganic Chemistry | 100 | 0.97 | 0.92 | 77 | 0.98 | 0.91 |
| Libr & Inf Sci | D-Lib Magazine | 100 | 0.85 | 0.79 | 17 | 0.74 | 0.69 |
| | Scientometrics | 100 | 0.92 | 0.72 | 37 | 0.92 | 0.63 |
| Polit Sci | Am J Political Sci | 100 | 0.94 | 0.86 | 57 | 0.94 | 0.90 |
| | Eur J Political Res | 100 | 0.91 | 0.90 | 36 | 0.84 | 0.81 |
| Virology | Journal of Virology | 100 | 0.78 | 0.84 | 87 | 0.75 | 0.83 |
| | PLoS Pathogens | 100 | 0.93 | 0.90 | 81 | 0.92 | 0.87 |

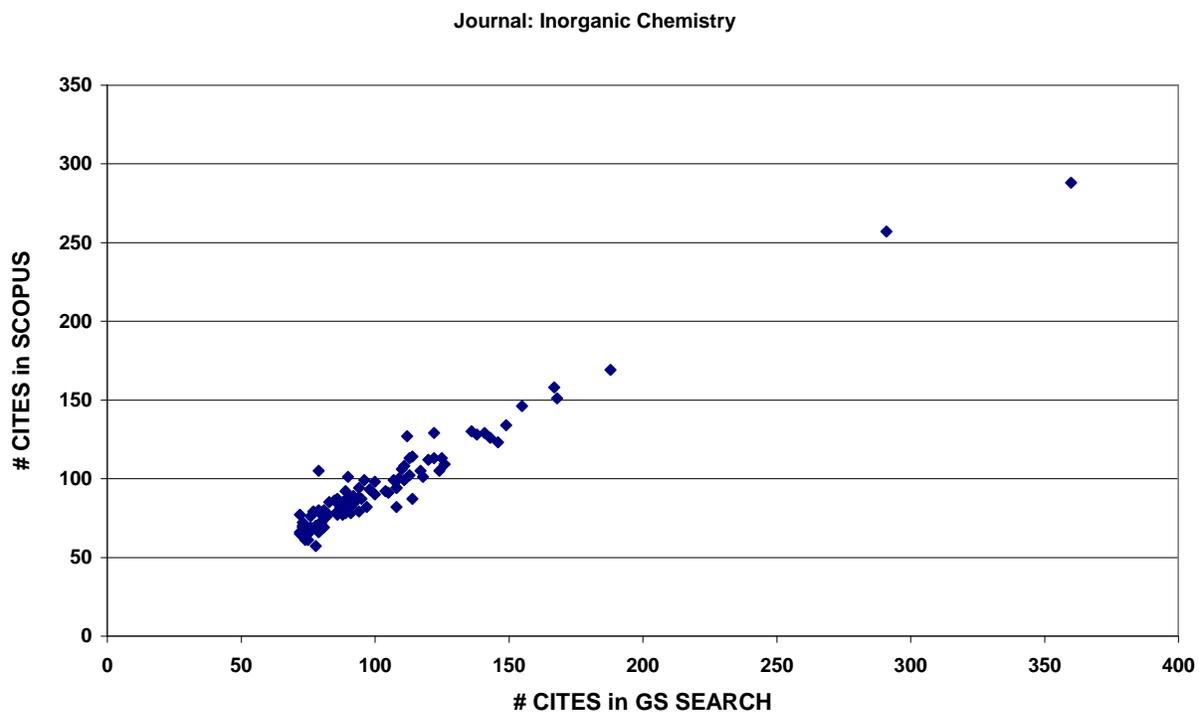

Figure 3. Scatterplot cites in Scopus against cites in Google Scholar Search per article in Inorganic Chemistry

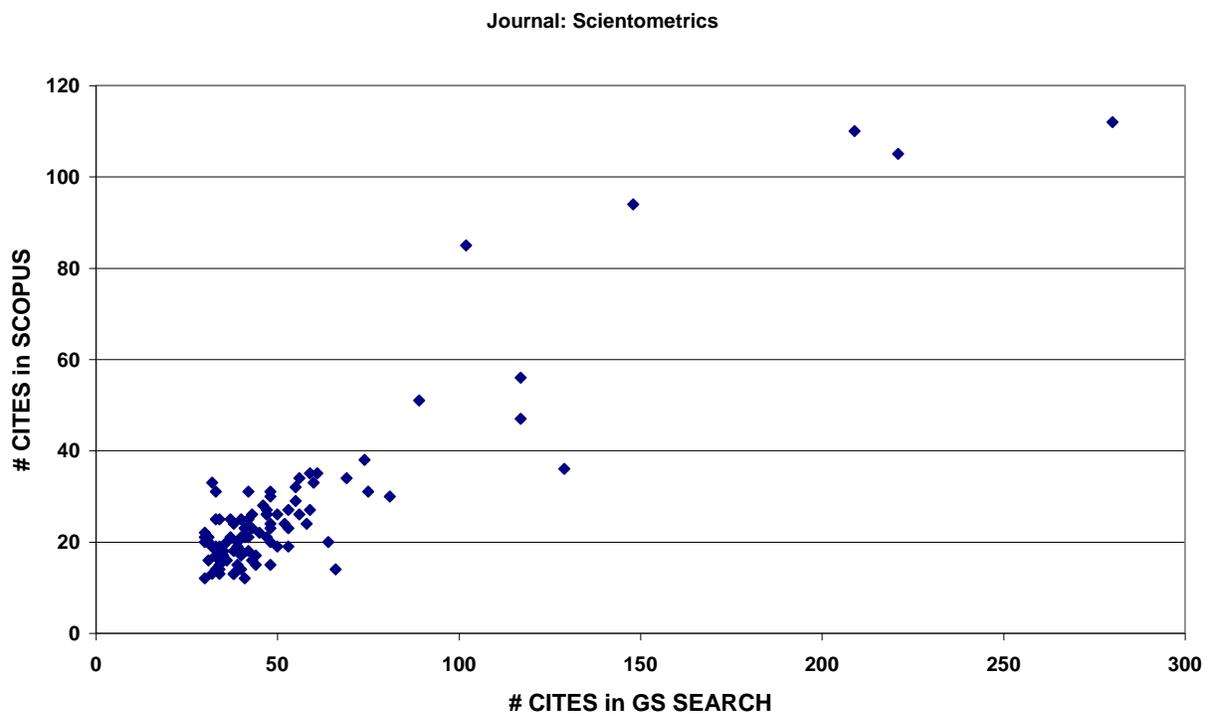

Figure 4. Scatterplot cites in Scopus against cites in Google Scholar Search per article in the journal Scientometrics

Focusing on the correlation between GS Search and Scopus counts, Table 10 shows that all Pearson coefficients are above 0.8, and for 8 out of 12 journals above 0.9. The Spearman coefficients are all above 0.7, and in three cases greater than or equal to 0.9. Table 11 suggests that the degree of correlation between Google Scholar and Scopus citation counts at the article level is independent of the volume of the GS Surplus. For instance, the two political science journals which according to Table 5 have GS/Scopus citation ratios around or above 3.0, have Pearson and Spearman coefficients that are very similar to those related to *Inorganic Chemistry*, and *PLoS Pathogens*, with citation ratios below 1.5 .

*3.6     Indexing speed*

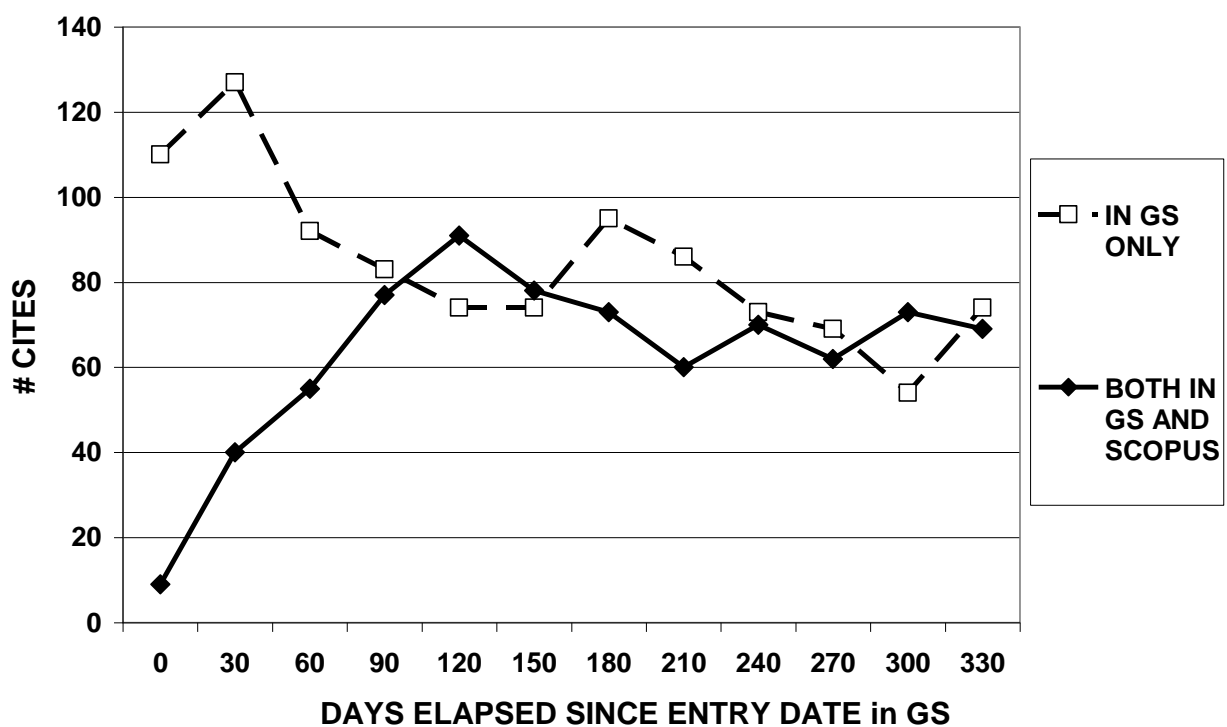

Figure 5. Number of citations in GS found and not found in Scopus as a function of the time elapsed since their entry date in GS. The horizontal axis indicates the time elapsed since the citations' entry date in GS, relative to the date of data collection in the study, expressed in time periods of 30 days (roughly speaking, one month). For instance, '0' means: the time elapsed is between 0 and 30 days; '30' means: between 30 and 60 days, and so forth.

Figure 5 relates to citing documents entered into GS during the 365 days prior to the date the data were collected in the current study. Figure 5 shows the absolute number of documents indexed in GS only, and the number indexed both in GS and in Scopus, as a function of the number of days elapsed since their entry date in GS. This figure clearly shows that the ratio of the two numbers changes as a function of the time elapsed since entry date. In fact, its value is above 10 in the first month, around 3.5 in the second, then further declines and seems to stabilize at a level of approximately 1.0 in the remaining part of the back-year. This value is equal to that one can obtain from the indicators for all target journals combined in Table 6 in Section 3.1. This outcome illustrates that, – at least for the journals in the study set, and at the end of July 2015, – only a small fraction of the citing documents entered in the preceding three months into Google Scholar was found in Scopus.

At least the following two factors may be responsible for this pattern: differences in coverage of indexed sources, and differences in indexing speed. To separate these two factors, and examine whether Google Scholar indexes documents faster than Scopus does, the categorization of citations presented in Table 7 in Section 3.2 was applied. As indicated in Section 3.2, a fraction of citations in journals indexed in Scopus but not found in Scopus are given in so called *articles in press*, the meta data of which are indexed in Scopus, but *not* their cited reference lists. The citations in Google Scholar contained in Scopus-indexed journals were subdivided into three subcategories: a) citations found in Scopus at the date of data collection; b) citations not found in Scopus but possibly included in source articles indexed in Scopus as article in press (AIP); c) citations not found in Scopus and probably not included in source articles indexed in Scopus as AIP. This sub-categorization can only be made if the source title in GS is available. In the data collected in the current study, this is only the case if the citation is found in Google Scholar Metrics.

While Figure 5 gives an impression on the absolute numbers of citations in the various sub-categories of citations, Figure 6 presents percentages relative to the total number of citations. It shows a breakdown of GS citations in Scopus-indexed journals into the three subcategories mentioned above, as a function of the time elapsed since citations' entry date in GS. The figure shows that in the set of citations entering GS in the second month before the date of data collection (days 30-60, indicated as '30' in Figure 6) and included in journals indexed in Scopus, only 52 per cent was actually found in Scopus at the download date. This percentage increases with increasing time elapsed since citations' entry date in GS, and reaches for citations included in the twelfth month prior to the date of data collection a value of almost 90 per cent. Still in the second month, 29 per cent was not found in Scopus but possibly included in reference lists of source articles indexed in Scopus as article in press (AIP). This percentage declines rapidly to the level of a few per cent at the end of the time period considered. The percentage of citations in source articles probably not indexed in Scopus as AIP declines as well, from 19 to 5 per cent.

These findings suggest that the indexing speed of Scopus-covered journals in GS is faster than that of the same journals in Scopus. The delay is largely, but not exclusively, caused by the fact that reference lists in articles in press are added with a delay into Scopus. Figure 6 shows that some 10 per cent of citing documents in Scopus-indexed journals are included in Scopus with a delay of more than one year compared to their entry date in GS. But here it must be underlined that among the latter documents some may be published in journals that were not yet indexed in Scopus in 2014 or earlier, as the criterion for being considered as a Scopus-indexed journal was that it is included in the list of active journals in June 2015.

Figure 6 shows the outcomes of a *synchronous* approach, as it analyzes a series of 'cohorts' of documents entering GS at various points in time during a time period of 12 months, and establishes whether they are indexed in Scopus at one fixed point in time, namely the date of download of the data. Assuming that this pattern is statistically similar to a *diachronous* pattern, in which one follows one fixed cohort of documents over a time period of 12 months, Figure 6 suggests that the median difference in delay between GS and Scopus of indexing documents in Scopus-covered journals is about 2 months, and the third quartile of this difference is about 4 months.

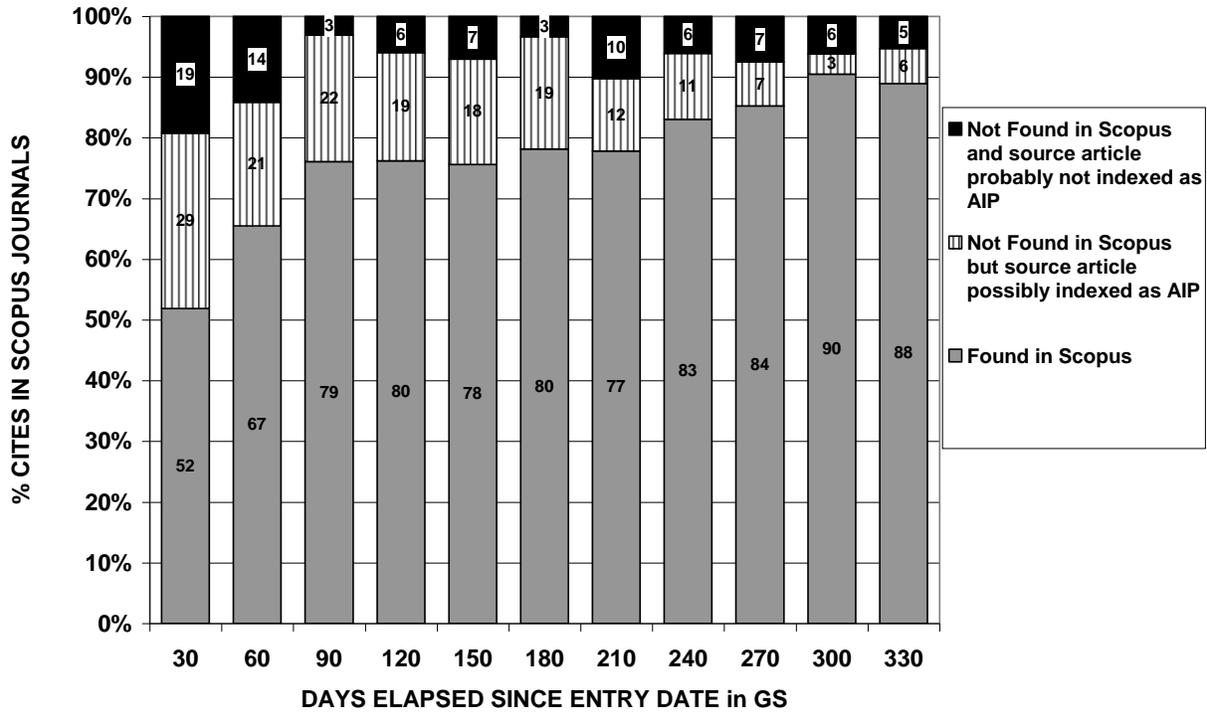

Figure 6. The percentage of citations in GS published in Scopus journals and found/not found in Scopus as a function of the time elapsed since the citations' entry date in GS. The horizontal axis indicates the time elapsed since the citations' entry date in GS, relative to the date of data collection in the study, expressed in time periods of 30 days (roughly speaking, one month). For instance, '30' means: the time elapsed is between 30 and 60 days; '60' means: between 60 and 90 days, and so forth. Contrary to Figure 5, there are no citations with an elapsed time between 0 and 30 days. This is because such citations could not be included in GS Metrics, as the latter database was 'frozen' in June 2015, while citation data in the study were collected in the last week of July 2015. Since the information about source titles was extracted from GS Metrics, they were not available in the current study.

# 4    Discussion and conclusions

## 4.1    Important general comments

In this section the main findings are summarized and discussed. They are grouped by the research questions addressed in Section 1. The summary aims to capture the essential features and tendencies; for details the reader is refereed to Sections 2 and 3. As a start, a series of important comments of a more general nature are made.

Firstly, Google Scholar and Scopus develop continuously; they are expanding their coverage, and further enhance data quality. The empirical results presented in this paper were based on data collected at the end of July 2015; hence, outcomes could be out-of-date already, or the effect of recent changes may not yet be fully visible. In their review (Orduna-Malea, Ayllón, Martín-Martín & López-Cózar, 2015), the authors conclude that Google Scholar has improved its coverage over the years. Other recent studies show significant improvement of Google Scholar coverage compared to its early years (Degraff et al. (2013) related to geosciences; Harzing (2014) in chemistry and physics; De Winter et al. (2014). Scopus has launched in 2014 a Book Citation Index Project, aiming to add around 75,000 books to Scopus by the end of 2015 (Meester, 2013). Since the data for the current study were collected in July 2015, the effects of this program can be expected to be only partially visible in the results.

Secondly, the results are based on articles published in 12 journals from six subject fields. They cannot be directly generalized. Hence, the study is exploratory, and hypothesis-generating rather than hypothesis-testing.  It indicates orders of magnitude, and differences among journals and subject fields. The main contribution of this study is its methodological design. It proposes a methodology for collecting and analyzing data from Google Scholar, and for comparing its performance with that of other databases.

## 4.2    Discussion per research question

### Source coverage

The study confirms findings by Kousha and Thelwall (2008) who observed that the amount of extra citation in GS is field-dependent, and largest in social sciences and in fields with a vast conference literature,. The low ratio of GS over Scopus citations obtained for European Journal of Inorganic Chemistry suggests that the data problems in GS discussed by Harzing (2014) are not yet fully solved. The most important sources indexed in Google Scholar but not in Scopus tend to come from Google Books, and from large disciplinary repositories and scholarly platforms. Sources in Scopus not indexed –or partially indexed – in Google Scholar are mainly books and Chinese journals.

The analysis of GS sources is complicated as there is no full list available of sources indexed for Google Scholar – criticized by Ortega (2014) and Orduna-Malea, Ayllón, Martín-Martín & López-Cózar (2015) and many other authors cited above – while on the other hand the information about Scopus coverage is comprehensive and of good quality.

Due to their variability, it is difficult to successfully match book and conference proceedings titles from different databases. Therefore, the lists of sources and web domains presented in Appendex 1 and classified as being found in one database but not in the other, may contain errors.

*The position of OA journals*

The observation that Open Access journals show a higher ratio of GS over Scopus citations than their non-OA counterparts is based on 3 cases only, and is obviously not statistically significant. But it at least shows that 'being OA' could be a significant factor explaining the value of this ratio. This hypothesis could be examined in a follow-up study analyzing a much larger sample of OA and non-OA journal pairs in a variety of subject fields. Such a study could be conducted along the lines proposed by Moed (2012), who claims that most citation studies on the effects of OA are biased to the extent that these are based on citation analyses carried out in a citation index with a selective coverage of the good, international journals. The use of a citation index with more comprehensive coverage, especially of the OA literatures, may reveal effects of OA upon citation impact that have been invisible in earlier studies.

*Citation impact of sources*

The citation analysis suggests that, in terms of Eugene Garfield's notions of a citation index, both databases cover a set of core sources in the fields studied. The surplus sources covered in GS but not in Scopus and vice versa tend to be sources with a more peripheral status in terms of citation impact.

*Statistical correlations*

The finding that citation counts at the article level in GS and Scopus show a strong (linear) statistical correlation suggests that the two databases are to some extent interchangeable, at least as far as citation counts are concerned of targets indexed in both. On the other hand, a non-negligible amount of the variation in GS counts is not explained by Scopus counts and vice versa. This means that other factors account for differences in the citation distributions in the two databases. Case studies of articles showing large discrepancies between GS and Scopus counts are a good first step.

*Indexing speed*

The median difference in delay between GS and Scopus of indexing documents in Scopus-covered journals is about 2 months. This finding suggests that the indexing speed of Scopus-covered journals in GS is faster than that of the same journals in Scopus. The delay is largely, but not exclusively, caused by the fact that reference lists in articles in press are added with a delay into Scopus

*The effect of duplicates on citation counts*

In Google Scholar different versions of an article are indexed, expressed in different documents, but in the data samples studied, citation counts are hardly affected by double counts. Duplicates in a strict sense, with identical meta data, are rare: 0.2 - 0.3 per cent for GS and 0.6 per cent for Scopus. Defining similarity in terms of a substantial overlap in document titles, author lists and publication years, the percentage of similar documents is around 2 per cent, and is lower for Scopus than it is for Google Scholar.

Even if two documents have identical titles and authors, the full texts, and especially the cited reference lists, may be different. Often a journal version has an extended bibliography compared to an earlier conference proceeding version. Hence, from the point of view of citation-based retrieval, a citation search of a document cited in the long but not in the short version does not necessarily retrieve both versions. The authors of the current paper are not aware on the basis of which criteria GS decides which reference list should be included in the preferred version.[1]

*Data quality and consistency*

As indicated in the literature review section, a series of studies has concluded that in many subject fields Google Scholar has a "better recall" but a "poor precision" (Wakimoto, 2014, on researchers in nephrology), or "strong coverage" but "somewhat unreliable" (Mingers and Lipitakis (2010) on business and management studies), showing the largest number of "inconsistencies" in content compared to Web of Science and Scopus (Adriaanse & Rensleigh (2013) on environmental sciences) and "questionable" in terms of accuracy and completeness (Bornmann et al., 2009, on citations to Zeritschrift fur Angewandte Chemie). Also Orduna-Malea, Ayllón, Martín-Martín & López-Cózar (2015) and Martin-Martin et al. (2015) concluded that GS's precision capabilities in terms of search are still lacking. The current study made a series of observations which point into the same direction

In any information system a trade-off has to be established between indexing speed and data accuracy. The current findings suggest that in GS this trade-off is made in favor of indexing speed. The question must be raised whether a database primarily aiming at high indexing speed provides an adequate basis for the calculation of accurate indicators of research performance and visibility, and whether, if one aims at producing such indicators, additional investments are needed to convert the database from a bibliographic to a bibliometric one.

*4.3    From an article-to-article index to a concept-to-concept index*

From a *theoretical* perspective, the issue should be raised on the basis of which criteria one may conclude that two documents are actually duplicates? For instance, if a document is published in a conference proceeding, and exactly the same document is later published in a journal, one may conclude that these two are genuine duplicates. Also, if the contents of the two versions are identical, but the latter is copy-edited, by the publisher, one can argue that the two versions are the same – at least from the point of view of their content. But if a first version of a document is published in ArXiv, and a revised version, based on peer review, is published in a journal, are these two versions duplicates? And what if the author lists are different, especially if the first author changes, or if authors are added or deleted?

---

[1] For instance, an article published by Kleinberg (1999) cites a paper published by Pinski and Narin (1976), but an earlier article by the same author (Kleinberg, 1998) with the same title does not. Scopus gives two citations to the 20 something overlapping references, but GS gives only one, since the versions are merged

These questions point towards an important difference between GS and Scopus: if one considers the publication of a published journal article as an endpoint of an article production process in which a series of subsequent draft versions are being generated, it can be maintained that Scopus reflects the endpoint of this process, and Google Scholar reflects not only the end point but also the intermediary stages of the publication process. In fact, according to the latter perspective, a journal article is not an endpoint but rather an intermediary result itself, and perhaps not always the most important one.

Following up on this perspective, a series of subsequent versions does not so much represent a particular article, but rather a *'concept'*, and shows how it is being developed over time (Moed and Visser, 2007). A typical example is a document series starting with a discussion paper posted on a personal website, followed by one or more conference presentations and publications in conference proceedings, and finally published in a peer reviewed journal. Bar-Ilan (2006) also discussed the issue of multiple publications of the same concept. She called them multiple manifestations of the same work or concept, following IFLA's Functional Requirements for Bibliographic Records (FRBR) (IFLA,1998).

The term concept is used in a broad sense, marks the key cognitive content of the series of documents, and points towards what it is that is being developed, for instance, a general notion, a specific hypothesis, a particular method. This opens the way to the view of measuring *concepts* rather than of *articles*, and of concept-to-concept rather than article-to-article citation index, in which document titles, bylines and even the full text of the various versions tend to be substantially similar, but not necessarily identical.

Adopting a more classical viewpoint, one could argue that what really counts is a final version, peer reviewed and hence meeting at least a set of professional standards, and registered in a global publication registry. Moreover, the final version embodies the earlier versions and represents a *synthesis* of the development. Its creation is made by the authors themselves, and is *not* left to the reader, and thus enhances the *efficiency* of the communication process. The sources of such final versions are most often journals, but, depending upon the subject field, may be (peer reviewed) books or conference proceedings as well.

Citation analysts are confronted with the question as to whether methods developed in citation indexes covering mainly peer reviewed journals (Science Citation Index, Web of Science, and also Scopus) can be directly implemented into an environment as Google Scholar. Would analyzing concept-to-concept citations *as if* they were article-to-article citations lead to invalid, distorted or non-interpretable results?

The empirical results presented above suggest that citation counts in GS are hardly affected by double counts. This statement relates to citations to *individual target articles*. Double counts *do* occur *at the level of target sources*, in the following way: if a document is, for instance, first published in ArXiv, and a next version later in a journal J, citations to the two versions are aggregated. In GS Metrics, in which ArXiv is included as a source, this document (assuming that its citation count exceed the h5 value of ArXiv and journal J) is listed *both* under ArXiv *and* under journal J, with *the same*, aggregate citation count.

This reveals that in a concept-to-concept index, in which concepts are spread over a series of documents, published in different sources, a construct like a "source" impact factor may not have much meaning anymore. Calculating such a metric is not so much the assessment of a source, but rather the assessment of a set of concepts spread over multiple documents. But one could also argue that it has always been this way, but that now, in a database like Google Scholar, harvesting seemingly scholarly documents from all over the internet, this phenomenon is being recorded, made more clearly visible, and more easily subject to further informetric study.

### 4.4    Future research directions

This work concentrated on the methodological aspects which were supported by an empirical study. Further empirical studies are warranted, e.g. to measure error rates, to verify the citations, to study further subject fields, non-English journals, lower ranked journals, OA versus non-OA journals, indexing speed and correlations.

### Acknowledgement

The authors are grateful to two anonymous referees for their valuable comments on an earlier version of this article.

## References

Bakkalbasi, N., Bauer, K., Glover, J., & Wang, L. (2006). Three options for citation tracking: Google Scholar, Scopus and Web of Science. *Biomedical Digital Libraries*, 3(7), doi: 10.1186/1742-5581-3-7.

Bar-Ilan, J. (2006). An ego-centric citation analysis of the works of Michael O. Rabin based on multiple citation indexes. Information *Processing and Management*, 42(6), 1553-1566.

Bar-Ilan, J. (2008). Which h-index? - A comparison of WoS, Scopus and Google Scholar. *Scientometrics*, 74(2), 257-271.

Bar-Ilan, J. (2010). Citations to the "Introduction to informetrics" indexed by WOS, Scopus and Google Scholar. *Scientometrics*, 82(3), 495-506.

Bornmann, L., Marx, W., Schier, H., Rahm, E., Thor, A., & Daniel, H. (2009). Convergent validity of bibliometric Google Scholar data in the field of chemistry - Citation counts for papers that were accepted by Angewandte Chemie International edition or rejected but published elsewhere, using Google Scholar, Science Citation Index, Scopus, and Chemical Abstracts. *Journal of Informetrics*, 3(1), 27-35.

Degraff, J. V., Degraff, N., & Romesburg, H. C. (2013). Literature searches with Google Scholar: Knowing what you are and are not getting. *GSA Today*, 23(10), 44-45.

De Groote, S. L., & Raszewski, R. (2012). Coverage of Google Scholar, Scopus, and Web of Science: A case study of the h-index in nursing. *Nursing Outlook*, 60(6), 391-400.

Delgado López-Cózar, E. D., Robinson-García, N., & Torres-Salinas, D. (2014). The Google Scholar experiment: How to index false papers and manipulate bibliometric indicators. *Journal of the Association for Information Science and Technology*, 65(3), 446-454.

de Winter, J. C. F., Zadpoor, A. A., & Dodou, D. (2014). The expansion of Google Scholar versus Web of Science: A longitudinal study. *Scientometrics*, 98(2), 1547-1565.

Haddaway, N. R., Collins, A. M., Coughlin, D., & Kirk, S. (2015). The Role of Google Scholar in Evidence Reviews and Its Applicability to Grey Literature Searching. *PloS One*, 10(9), e0138237.

Harzing, A-W. (2014). A longitudinal study of Google Scholar coverage between 2012 and 2013. *Scientometrics*, 98(1), 565-575.

Harzing, A-W. (2013). A preliminary test of Google Scholar as a source for citation data: A longitudinal study of Nobel prize winners. *Scientometrics*, 94(3), 1057-1075.

Harzing, A-W., & Alakangas, S. (2016). Google Scholar, Scopus and the Web of Science: a longitudinal and cross-disciplinary comparison. *Scientometrics,* 106(2), 787-804.

IFLA (1998). Functional requirements for bibliographic records. Final report. UBCIM publications – New series, Vol. 19. Available from http://www.ifla.org/VII/s13/frbr/frbr.pdf.

Jacsó, P. (2005). Google Scholar: The pros and the cons. *Online Information Review*, 29(2), 208-214.

Jamali, H. R., & Nabavi, M. (2015). Open access and sources of full-text articles in Google Scholar in different subject fields. Scientometrics, 105(3), 1635-1651.

Kleinberg, J. M. (1998). Authoritative sources in a hyperlinked environment. December 1997 SODA '98: Proceedings of the Ninth Annual ACM-SIAM Symposium on Discrete Algorithms.

Kleinberg, J. M. (1999). Authoritative sources in a hyperlinked environment. Journal of the ACM (JACM), 46, 604-632.

Kousha, K., & Thelwall, M. (2007). Google Scholar citations and google Web/URL citations: A multi-discipline exploratory analysis. *Journal of the American Society for Information Science and Technology*, 58(7), 1055-1065.

Kousha, K., & Thelwall, M. (2008). Sources of Google Scholar citations outside the Science Citation Index: A comparison between four science disciplines. *Scientometrics*, 74(2), 273-294.

Kulkarni, A. V., Aziz, B., Shams, I., & Busse, J. W. (2009). Comparisons of citations in Web of Science, Scopus, and Google Scholar for articles published in general medical journals. *JAMA - Journal of the American Medical Association*, 302(10), 1092-1096.

Levine-Clark, M., & Gil, E. L. (2009). A comparative citation analysis of Web of science, Scopus, and Google Scholar. *Journal of Business and Finance Librarianship*, 14(1), 32-46.

Mayr, P., & Walter, A. (2007). An exploratory study of Google Scholar. *Online Information Review*, 31(6), 814-830.

Meester, W. (2013). Towards a comprehensive citation index for the Arts & Humanities. Research Trends, Issue 32, March 2013. http://www.researchtrends.com/issue-32-march-2013/towards-a-comprehensive-citation-index-for-the-arts-humanities/.

Meho, L. I., & Yang, K. (2007). Impact of data sources on citation counts and rankings of LIS faculty: Web of Science versus Scopus and Google Scholar. *Journal of the American Society for Information Science and Technology*, 58(13), 2105-2125.

Mingers, J., & Lipitakis, E. A. E. C. G. (2010). Counting the citations: A comparison of Web of Science and Google Scholar in the field of business and management. *Scientometrics*, 85(2), 613-625.

Moed, H.F. (2012). Does open access publishing increase citation or download rates? http://www.researchtrends.com/issue28-may-2012/does-open-access-publishing-increase-citation-or-download-rates/. Retrieved, 3 December 2015.

Neuhaus, C., Neuhaus, E., Asher, A., & Wrede, C. (2006). The depth and breadth of Google Scholar: An empirical study. *Portal*, 6(2), 127-141.

Orduna-Malea, E., Ayllón, J. M., Martín-Martín, A., & López-Cózar, E. D. (2015). Methods for estimating the size of Google Scholar. *Scientometrics*, 104(3), 931-949.

Ortega, J. L. (2014). Academic search engines: A quantitative outlook. Elsevier.

Pinski, G., and Narin, F. (1976). Citation influence for journal aggregates of scientific publications: theory, with application to the literature of physics. Information Processing and Management, 12, 297–312.

Pitol, S.P., De Groote, S.L . (2014). Google Scholar versions: do more versions of an article mean greater impact? Library Hi Tech, 32(4), 594-611.

Scopus Journal Title List (2015). Available at: https://www.elsevier.com/solutions/scopus/content. Last accessed: 10 December 2015.

Valderrama-Zurián, J. C., Aguilar-Moya, R., Melero-Fuentes, D., & Aleixandre-Benavent, R. (2015). A systematic analysis of duplicate records in Scopus. Journal of Informetrics, 9(3), 570-576.

Wildgaard, L. (2015). A comparison of 17 author-level bibliometric indicators for researchers in Astronomy, Environmental Science, Philosophy and Public Health in Web of Science and Google Scholar. *Scientometrics*, 104(3), 873-906.

**Appendix**

*A1. Match-merge*

The total number of documents in the first data row includes both target articles and citing documents. A large degree of similarity was defined as follows: document titles should have at least three title words and at least 50 per cent of the words in each title in common; author lists should have at least one author last name in common, and publication years should differ at most 2 years. In the matching process between GS and Scopus based on the four match-keys in Table 4, only the matches of two documents being identical or showing a large degree of similarity were selected.

Table 4: Number of pairs containing identical or similar documents

| Dataset | GS Search | | GS Metrics | | Scopus | |
|---|---|---|---|---|---|---|
| | N | % | N | % | N | % |
| Total # docs | 10,908 | | 7,068 | | 5,918 | |
| Total duplicate pairs | 430 | 3.9 | 366 | 5.2 | 119 | 2.0 |
| identical | 26 | 0.2 | 18 | 0.3 | 34 | 0.6 |
| Large degree of similarity (but not identical) | 205 | 1.9 | 174 | 2.5 | 60 | 1.0 |

Apart from the fact that the measurement of document similarity is based on meta data only and ignores the full text, and from the conceptual issue of when two documents are genuinely identical, an issue that will be discussed in the concluding section, the definition applied here has a certain degree of arbitrariness. It can be concluded that duplicates with identical meta data are rare: 0.2 - 0.3 per cent for GS and 0.6 for Scopus. Defining similarity in terms of a substantial overlap in document titles, author lists and publication years, the percentage of similar documents is a few per cent., and is larger for GS than it is for Scopus. The percentage of duplicate pairs generated by a particular match-key can perhaps be interpreted as an upper bound. It is 4-5 per cent for Google Scholar and 2 per cent for Scopus. In the analyses presented below, duplicate documents showing a large degree of similarity or being identical were deleted from the data files.

*A2 – Sources and web domains*

Table A2 present the ten most frequently appearing sources or web links from each distribution. The term "source" is used in a broad sense, and does not only include journal titles, book or proceedings titles, but also the names of institutions hosting the repositories in or via which the full texts of the documents are available. It must be noted that they are often incomplete and not standardized, so that it can not be excluded that occasionally entries were misclassified. This is especially true for books and conference proceedings volumes, the titles of which are not as well standardized as those of journals. But such errors do not affect the overall pattern in the data.

Table A2. The 10 most frequently occurring web links and sources in Google Scholar not found in Scopus and vice versa

| Data field | # Citing Docs | Comments |
|---|---|---|
| The 10 most frequently occurring *web links* of citing documents in sources not indexed in Scopus | | |
| books.google.com | 156 | Google's Book Index |
| Springer | 140 | Monographs, book chapters and proceedings papers published by Springer |
| papers.ssrn.com | 93 | Documents posted in the Social Sciences Research Network |
| researchgate.net | 86 | Social networking site for scientists and researchers to share papers, communicate and find collaborators |
| dl.acm.org | 63 | ACM Digital Library containing full texts of all articles published by ACM |
| arxiv.org | 54 | A repository of freely available e-prints of scientific papers in physics, mathematics, computer science and other fields |
| aclweb.org | 53 | Website of the Association for Computational Linguistics |
| anthology.aclweb.org | 39 | Digital Archive of research papers in Computational Linguistics |
| Wiley Online Library | 38 | Monographs, book chapters and proceedings papers published by Wiley |
| ieeexplore.ieee.org | 36 | Provides abstracts and full-text articles on computer science, electrical engineering and electronics, mainly published by IEEE and IET |
| The 10 most frequently occurring *sources* of citing documents indexed in GS Metrics but *not* published in sources indexed in Scopus | | |
| Source in GS Metrics | # Citing Docs | Comments |
| arXiv preprint arXiv: | 53 | A repository of freely available e-prints of scientific papers in physics, mathematics, computer science and other fields |

| Available at SSRN | 36 | SSRN is Social Sciences Research Network |
|---|---|---|
| APSA | 22 | American Political Science Association |
| Palgrave Macmillan | 16 | Book publisher |
| Cambridge University Press | 12 | Book publisher |
| Proceedings of the 13th Annual Meeting of the Special Interest Group on ... | 11 | Incomplete title. Manual check shows that the full title is: SIGDIAL '12 Proceedings of the 13th Annual Meeting of the Special Interest Group on Discourse and Dialogue |
| ACL | 8 | Association for Computational Linguistics |
| Data-Driven Methods for Adaptive Spoken Dialogue Systems | 7 | Title of a book published by Springer in 2012 |
| Oxford University Press | 7 | Book publisher |
| Transactions of the Association for Computational Linguistics | 7 | Publication of the ACL |

The 10 most frequently occurring *sources* of citing documents indexed in Scopus not found or only partially covered in Google Scholar

| Progress in Chemistry | 4 | This journal published by the Chinese Academy of Sciences is only partially covered in GS |
|---|---|---|
| Revista Brasileira de Politica Internacional | 4 | Not found as source in GS |
| Acta Chimica Sinica | 3 | This Chinese journal is only partially covered in GS |
| Global Political Economy: Contemporary Theories, Second Edition | 3 | Book title |
| Gongneng Cailiao/Journal of Functional Materials | 3 | Chinese journal, not found as source is GS |
| Millennium: Journal of International Studies | 3 | Not found as source is GS |
| Rural Policy Implementation in Contemporary China: New Socialist Countryside | 3 | Book title |
| Alternatives | 2 | Not found as source is GS |
| American Bee Journal | 2 | Not found as source is GS |
| Beijing Daxue Xuebao (Ziran Kexue Ban)/Acta Scientiarum Naturalium Universitatis Pekinensis | 2 | Chinese journal, not found as source is GS |